# Quadrupolar Susceptibility and Magnetic Phase Diagram of PrNi$_2$Cd$_{20}$ with Non-Kramers Doublet Ground State

Tatsuya Yanagisawa[1*], Hiroyuki Hidaka[1], Hiroshi Amitsuka[1], Shintaro Nakamura[2], Satoshi Awaji[2], Elizabeth L. Green[3], Sergei Zherlitsyn[3], Joachim Wosnitza[3], Duygu Yazici[4], Benjamin. D. White[5], and M. Brian Maple[4]

[1] *Graduate School of Science, Hokkaido University, Sapporo 060-0810, Japan*
[2] *High Field Laboratory for Superconducting Materials, Institute for Material Research, Tohoku University, Sendai 980-8577, Japan*
[3] *Hochfeld-Magnetlabor Dresden (HLD-EMFL) and Würzburg-Dresden Cluster of Excellence ct.qmat, Helmholtz-Zentrum Dresden-Rossendorf, 01328 Dresden, Germany*
[4] *Department of Physics, University of California, San Diego, La Jolla, California 92093, USA*
[5] *Central Washington University, Department of Physics, Ellensburg, Washington 98926-7501, USA*

**ABSTRACT**

In this study, ultrasonic measurements were performed on a single crystal of cubic PrNi$_2$Cd$_{20}$, down to a temperature of 0.02 K, to investigate the crystalline electric field ground state and search for possible phase transitions at low temperatures. The elastic constant $(C_{11}-C_{12})/2$, which is related to the $\Gamma_3$-symmetry quadrupolar response, exhibits the Curie-type softening at temperatures below ~30 K, which indicates that the present system has a $\Gamma_3$ non-Kramers doublet ground state. A leveling-off of the elastic response appears below ~0.1 K toward the lowest temperatures, which implies the presence of level splitting owing to a long-range order in a finite-volume fraction associated with $\Gamma_3$-symmetry multipoles. A magnetic field–temperature phase diagram of the present compound is constructed up to 28 T for $H \parallel [110]$. A clear acoustic de Haas–van Alphen signal and a possible magnetic-field-induced phase transition at $H$ ~26 T are also detected by high-magnetic-field measurements.

## Introduction

In cubic Pr-based non-Kramers doublet systems, several unusual phenomena have been predicted theoretically and observed experimentally, such as the localized higher-rank multipolar order of a non-Kramers $\Gamma_3$ doublet ground state [1], vibronic states [2], coexistence of quadrupolar order and superconductivity [3], and a quadrupolar Kondo effect (QKE) caused by hybridization between conduction electrons and local $f$-electrons ($c$-$f$ hybridization) [4]. For example, PrInAg$_2$ [5, 6], PrPb$_3$ [7-9], PrMg$_3$ [10], PrV$_2$Al$_{20}$ [11], PrTi$_2$Al$_{20}$ [12], PrIr$_2$Zn$_{20}$ [13], and related compounds have been examined as candidate materials for evidencing these intriguing phenomena.

In the present study, we focused on PrNi$_2$Cd$_{20}$ with a cubic CeCr$_2$Al$_{20}$-type crystal structure in which the Pr and Ni crystallographic sites have cubic $T_d$ ($4\bar{3}m$) and trigonal $D_{3d}$ ($\bar{3}m$) point-group symmetries, respectively [14]. The Cd ions are located in three distinct crystallographic sites (16$c$ site: $\bar{3}m$, 48$f$ site: $mm$, and 96$g$ site: $m$) consisting of the polyhedra displayed in Fig. 1. [15]

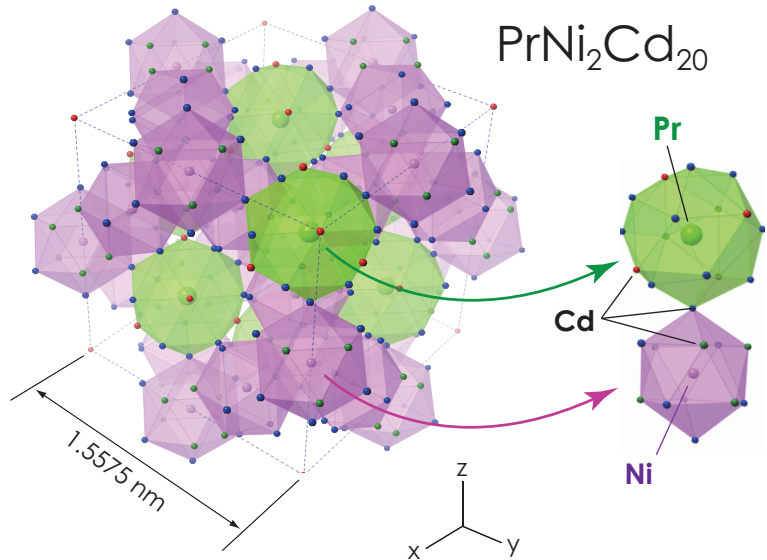

**Figure. 1**   Crystal structure of PrNi$_2$Cd$_{20}$.

In previous works, PrNi$_2$Cd$_{20}$ has been found to grow as a single crystal by the Cd self-flux method [17]. The lattice constant $a$ = 1.5575 nm is larger than that of PrIr$_2$Zn$_{20}$ ($a$ = 1.4273 nm), PrV$_2$Al$_{20}$ ($a$ = 1.4591 nm), and PrTi$_2$Al$_{20}$ ($a$ = 1.4723 nm), where the ionic radii, Cd$^{2+}$ > Zn$^{2+}$ > Al$^{3+}$. The relatively large Pr-Pr separation (of the order of 0.674 nm) in PrNi$_2$Cd$_{20}$ cause a moderate Ruderman–Kittel–Kasuya–Yosida (RKKY) interaction and weaker $c$-$f$ hybridization [16]. Therefore, PrNi$_2$Cd$_{20}$ can be considered as a good candidate to study a more localized Pr system as a reference to compare with the sister compounds PrTi$_2$Al$_{20}$ and PrIr$_2$Zn$_{20}$, which have stronger $c$-$f$ hybridization and possibly demonstrate the QKE. The specific-heat data for the PrNi$_2$Cd$_{20}$ show no evidence for magnetic order down to 2 K and indicate a $\Gamma_3$-doublet or a $\Gamma_1$-singlet crystalline electric field (CEF) ground state [17]. The magnetic entropy estimated from the specific heat reaches $R\ln 2$ at ~7 K, exceeding the fractional residual entropy $1/2 R\ln 2$, which was predicted from the theoretical QKE model by Cox [18]. Therefore, the QKE model might be ruled out for the present system. In that case, the Kondo effect will not last up to the strong-coupling regime for overcoming the RKKY interaction. Thus, $\Gamma_3$-type local quadrupolar order is expected for the low-temperature phase transition to release the residual $R\ln 2$ entropy when the temperature approaches absolute zero. As no magnetic phase transition has been observedusing magnetization measurements down to 0.5 K under magnetic fields up to 7 T so far [17], further investigations at very low temperatures and higher magnetic fields are required to understand the ground state of the 4f-electron system in the present compound.

## Experimental Details

In the present study, we have performed ultrasonic measurements on a single crystal of $PrNi_2Cd_{20}$ at temperatures as low as 0.02 K to investigate the CEF ground state and search for possible phase transitions. Single-crystalline $PrNi_2Cd_{20}$ samples were grown at UC San Diego using the Cd self-flux method. The orientation of the investigated single crystal was determined by the Laue X-ray method. As the present as-grown crystal contains several internal cavities, which cause undesired accidental reflections of the ultrasonic wave, careful sample cutting, and polishing have been conducted under the microscope to avoid these cavities. The sample dimensions are $2.525 \times 4.0 \times 2.5$ mm$^3$ for [110]-[1$\bar{1}$0]-[001]. An absolute value of the elastic constant $(C_{11}-C_{12})/2 = 2.047$ J/m$^3$ is estimated from the density $\rho = 8.814$ g/cm$^3$ and the absolute value of the transverse sound velocity $v = 1524$ m/s at 300 K. Ultrasonic measurements were carried out utilizing the pulse-echo method using $LiNbO_3$ transducers with a thickness of 100 μm. The sound velocity and ultrasonic attenuation were measured using the phase-comparative method. [19]

## Results and Discussion

Figure 2 shows the elastic constant $(C_{11}-C_{12})/2$ of $PrNi_2Cd_{20}$ as a function of temperature. The $(C_{11}-C_{12})/2$ mode, which is observed by a transverse ultrasonic wave propagating along the $\boldsymbol{k} \parallel [110]$ axis with polarization along the $\boldsymbol{u} \parallel [1\bar{1}0]$ axis, exhibits softening at temperatures below 30 K and then levels off below ~0.1 K toward lowest temperatures. It should be noted that no signature of the ultrasonic dispersion owing to the thermally activated rattling motion of atoms, which is found in $PrRh_2Zn_{20}$, $PrIr_2Zn_{20}$, some rare-earth-based clathrates, and filled-skutterudite compounds, is observed in the present ultrasonic measurements [20-23]. The quadrupolar susceptibility [19] of $O_2^2$ (= $J_x^2 - J_y^2$) for the $Pr^{3+}$ ($J = 4$) ion with following CEF models are shown in Fig.2; a $\Gamma_3$-doublet ground state and a $\Gamma_4$-triplet excited state at ~10.0 K ($x = +0.5$, $W = -0.4$ K, shown as green solid line in Fig. 2) and also a $\Gamma_3$-doublet ground state and a $\Gamma_5$-triplet excited state at ~10.5 K ($x = -0.5$, $W = -0.39$ K, shown as the blue dashed line). Both CEF models reproduce well with the elastic softening of $(C_{11}-C_{12})/2$ down to ~0.1 K, where the elastic constant $(C_{11}-C_{12})/2$ levels off toward absolute zero, deviating from the calculated quadrupolar susceptibilities. Here, the parameters $x$ and $W$ indicate two independent CEF parameters defined in Ref. [24]. On the other hand, another CEF model with a $\Gamma_1$-singlet ground state and a $\Gamma_5$-triplet excited state at 5.5 K ($x = +0.4$, $W = +0.38$ K, shown as the purple dotted line) cannot reproduce the low-temperature softening. Such a clear contrast indicates that the present system has at least a non-Kramers doublet ground state rather than the singlet–triplet pseudo-quartet state. The obtained coupling constant of the quadrupole-strain interaction is $|g_{\Gamma3}|$ = 13.5–14.0 K, which is much smaller than that of $|g_{\Gamma3}|$ = 920.8 K for $PrRh_2Zn_{20}$ and comparable to $|g_{\Gamma3}|$ = 30.9 K for $PrIr_2Zn_{20}$ [22]. A tiny and negative quadrupolar–quadrupolar intersite interaction coefficient $g'_{\Gamma3}$ = −3.5 to −4.0 mK is also obtained, which indicates the presence of a weak antiferroquadrupolar interaction that is much smaller than $g'_{\Gamma3}$ = −2.413 K for $PrRh_2Zn_{20}$ and $g'_{\Gamma3}$ = −0.13 K for $PrIr_2Zn_{20}$. Note that the abovementioned CEF models with $\Gamma_3$ non-Kramers doublet ground states reproduce the magnetic susceptibility well down to 2 K and are consistent with the Schottky peak that appears in the specific heat at approximately 5 K [17]. From the zero-magnetic-field data, however, we could not distinguish the true CEF-level scheme of the present compound, i.e., whether the system has a $\Gamma_3$–$\Gamma_4$ or the $\Gamma_3$–$\Gamma_5$ ground state. It should also be noted that the sister compound $PrIr_2Zn_{20}$ has the $\Gamma_3$–$\Gamma_4$ (26 K) ground state [25].

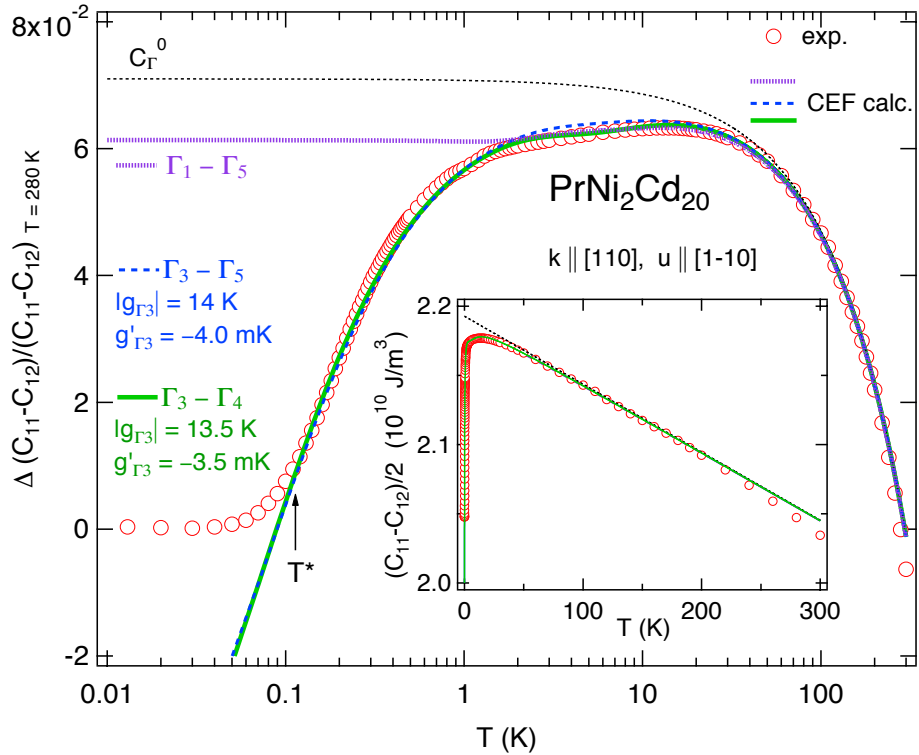

**Figure. 2** Relative change of the elastic constant $(C_{11}-C_{12})/2$ of $PrNi_2Cd_{20}$ vs. temperature. Purple dotted, blue dashed, and green solid lines represent the calculated quadrupolar susceptibilities for CEF ground-state models—$\Gamma_1$–$\Gamma_5$ (5.5 K), $\Gamma_3$–$\Gamma_5$ (10.5 K), and $\Gamma_3$–$\Gamma_4$ (10.0 K), respectively. Black dashed line corresponds to a phonon background used for the fits. The arrow indicates the temperature $T^*$, where the elastic constant deviates from the fit. The inset represents the data displayed as the absolute value vs linear temperature scale.

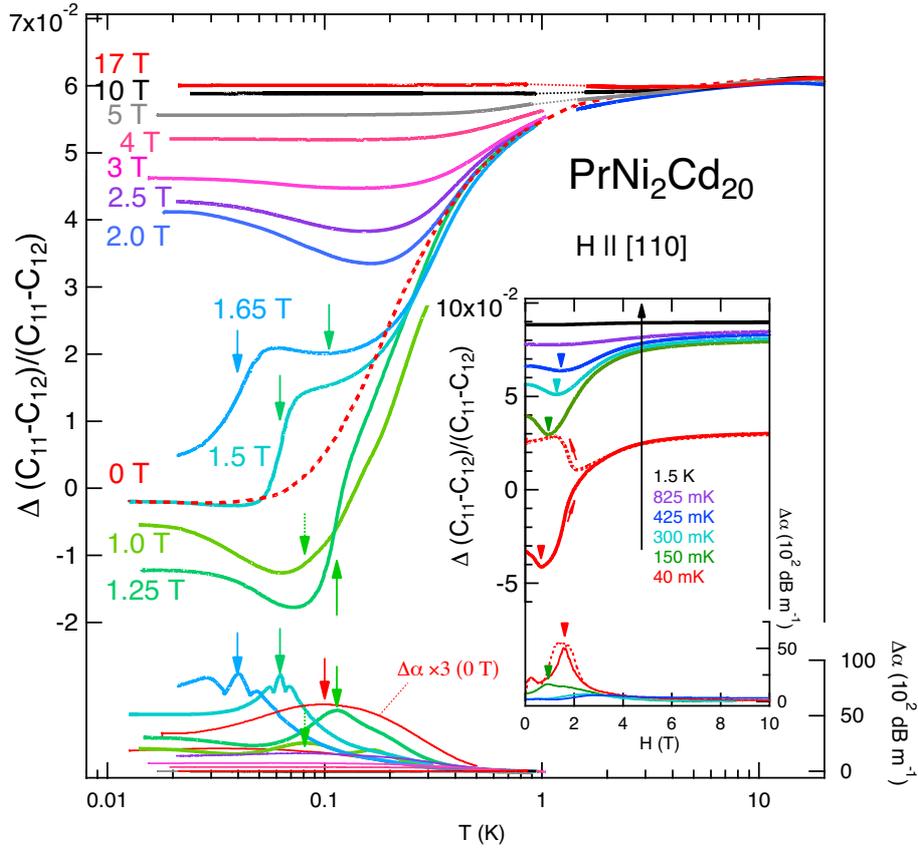

**Figure. 3** (Upper panel) Temperature dependence of $(C_{11}-C_{12})/2$ of $PrNi_2Cd_{20}$ shown as relative change, at low-temperatures below 20 K and under applied selected magnetic fields. (Lower panel) Ultrasonic attenuation coefficient $\Delta\alpha$ vs temperature. (Inset) $(C_{11}-C_{12})/2$ as a function of magnetic field at various fixed temperatures between 40 mK and 1.5 K.

The leveling-off feature of $(C_{11}-C_{12})/2$ below ~0.1 K implies a tiny level splitting of the $\Gamma_3$ doublet caused by a short-range order associated with the $\Gamma_3$-symmetry multipoles and/or the presence of a finite-volume fraction of disordered Pr sites as already discussed in Ref. [26]. The estimated energy scale of the tiny level splitting of the $\Gamma_3$ doublet is $\Delta \sim$ 48 mK, which can be mimicked by adding a small tetragonal distortion of $B_2^0 = 3$ mK in the cubic CEF Hamiltonian, and being considerably smaller than the larger CEF splitting of the first excited triplet of 5.5-10.5 K. This tiny level splitting reproduces well the leveling-off feature below ~0.1 K.

The main panel of Fig. 3 shows the relative change of the elastic constant $(C_{11}-C_{12})/2$ as a function of temperature at fixed magnetic fields up to 17 T. The elastic softening below 30 K is suppressed by applying a magnetic field sincethe nonmagnetic Kramers doublet obeys Zeeman splitting and mixing with the wave function from the first excited triplet. Intriguingly, the leveling-off feature at 0 T changes to a clear shoulder-like anomaly between 1.0 and 1.65 T; then the clear anomaly disappears above 2.0 T. In the lower panel of Fig. 3, the ultrasonic attenuation is shown using the same color scheme. The temperature dependence of the ultrasonic attenuation at zero magnetic field (red solid line) shows a very broad local maximum at approximately 0.1 K at 0 T. This peak in ultrasonic attenuation develops when magnetic fields of 1.0–1.65 T are applied at the same temperatures where the shoulder-like anomaly is observed in the elastic constant. In the inset of Fig. 3, the elastic constant $(C_{11}-C_{12})/2$ is shown as a function of magnetic field at various temperatures between 40 mK and 1.5 K. A clear minimum appears at $H \sim 1$–2 T. In the lowest temperature data at 40 mK, a pronounced hysteresis in ramping up and down of the magnetic field is observed below $H = 3$ T. These results, observed under applied magnetic fields, imply that there is a magnetic-field-induced long-range order in the narrow region between ~1 and 2 T for $H \parallel [110]$ below 150 mK.

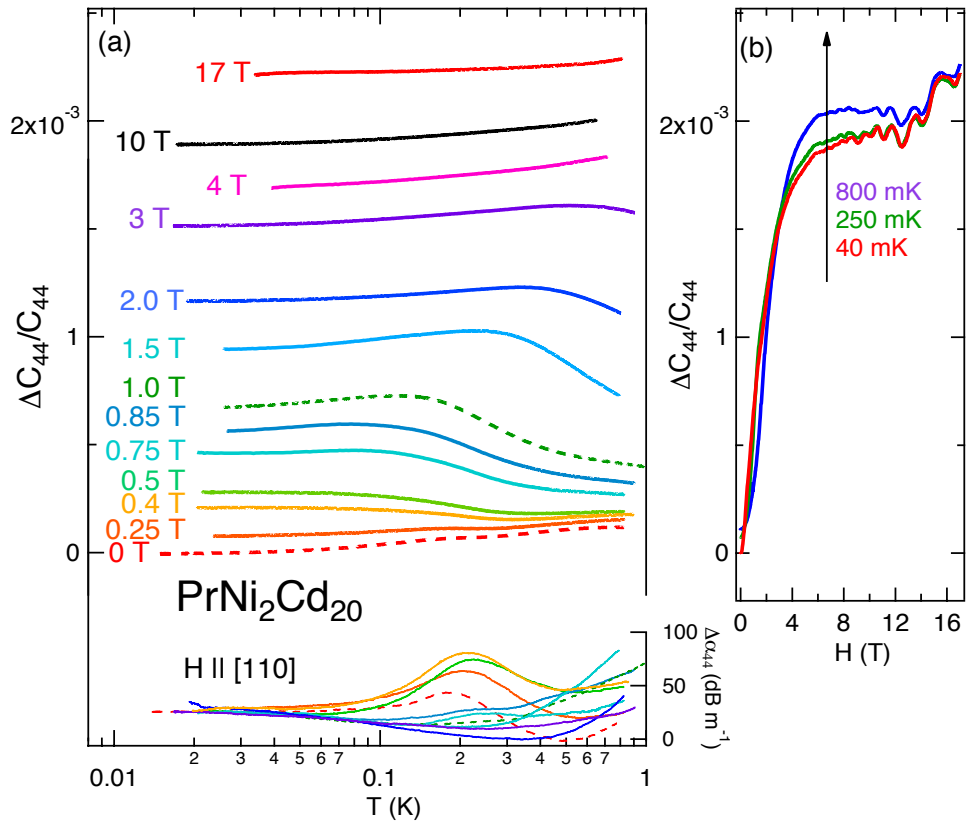

**Figure. 4** (a) (Upper panel) Temperature dependence of $C_{44}$ of PrNi$_2$Cd$_{20}$ shown as relative change, at selected magnetic fields for the low-temperature region, below 1 K. The data are vertically shifted to facilitate the display of the temperature and magnetic field variations. (Lower panel) Ultrasonic attenuation coefficient $\Delta\alpha_{44}$ vs temperature. (b) $C_{44}$ as a function of magnetic fields at various fixed temperatures.

To verify the possible phase transition in the present compound, we performed specific-heat measurements down to 80 mK (not shown) using a tiny piece of the crystal cleaved from the surface of the present single crystal sample. However, no evident anomaly is observed in the specific heat data. Instead, the specific heat data shows a monotonic increase, probably because of a nuclear Schottky contribution. From these discrepancies between the elastic response and the specific heat, it can be considered that there is no 'bulk' thermodynamic phase transition. An explanation for these discrepancies is the existence of possible inhomogeneities in the sample. If there is a long-range order in a finite-volume fraction in the present compound, the anomaly related to the region of long-range order will not be detected in specific-heat measurements using a tiny sample. As the present ultrasonic measurements were performed on a relatively large piece of the sample cleaved from the same single crystal that probably contains inhomogeneous regions, a local long-range order will be observed as an elastic anomaly in the bulk-sensitive methods. From the localized 4f-electron picture, another possibility to understand why the long-range order, collapsed in zero field, appears only in the lower magnetic-field range could be due to the tiny level splitting of the $\Gamma_3$ doublet as described above. Here, the level splitting of the $\Gamma_3$ doublet becomes a pseudo doublet by applying a weak magnetic field of ~1 T. We need further investigations to check the possible sample-quality dependence of the magnetic field-temperature (*H-T*) phase diagram as shown later in Fig. 7. Another possibility, which has not been considered in the present work, is the effect of nuclear hyperfine coupling between the Pr-nuclear spins and 4*f*-electron system, which had been discussed for Pr-based filled-skutterudite compounds [33].

By observing different ultrasonic modes, we can investigate the quadrupolar response with different symmetries. Figure 4 (a) shows the relative change of the elastic constant $C_{44}$ as a function of temperature at fixed magnetic fields up to 17 T. The $C_{44}$ mode is observed using a transverse ultrasonic wave propagating along the ***k*** || [110] axis with polarization along the ***u*** || [001] axis, related to the $\Gamma_5$ symmetry quadrupolar response. As the non-Kramers $\Gamma_3$ doublet does not have $\Gamma_5$ quadrupole degrees of freedom, no response is expected. Indeed, the change in the elastic constant $\Delta C_{44}/C_{44} \sim 1 \times 10^{-4}$ for 0 T is much smaller than that of $\sim 5 \times 10^{-2}$ in $(C_{11}-C_{12})/2$. This fact clearly indicates that these transverse modes do not mix and are also well distinguished from each other under the present measurement conditions. We can conclude that the CEF ground state of the present system is the $\Gamma_3$ doublet. $C_{44}$ also shows no clear elastic anomaly in the present temperature and magnetic-field region, except for a broad maximum around 0.2 K between $H$ = 0 and 1 T in the ultrasonic attenuation. The origin of the ultrasonic attenuation peak is unclear, and we need further investigations, such as the frequency dependence of $C_{44}$, in order to clarify this issue. Figure 4(b) represents the elastic constant $C_{44}$ as a function of magnetic field at 40, 250, and 800 mK. No minimum appears at approximately 1–2 T, in contrast to the elastic constant $(C_{11}-C_{12})/2$ (inset of Fig. 3). This contrast of the elastic response in magnetic field implies that the order parameter of the magnetic-field-induced long-range order is related to electric quadrupoles $O_2^0$ and $O_2^2$ or the magnetic octupole $T_{xyz}$ in the $\Gamma_3$-doublet state. In the high-magnetic-field region, $C_{44}$ shows clear de Haas–van Alphen (dHvA) oscillations.

In order to investigate the elastic constant $(C_{11}-C_{12})/2$ in higher-magnetic-field regions, we used the hybrid-magnet 28T-CHM at IMR, Tohoku University, combined with a $^3$He-$^4$He dilution refrigerator.

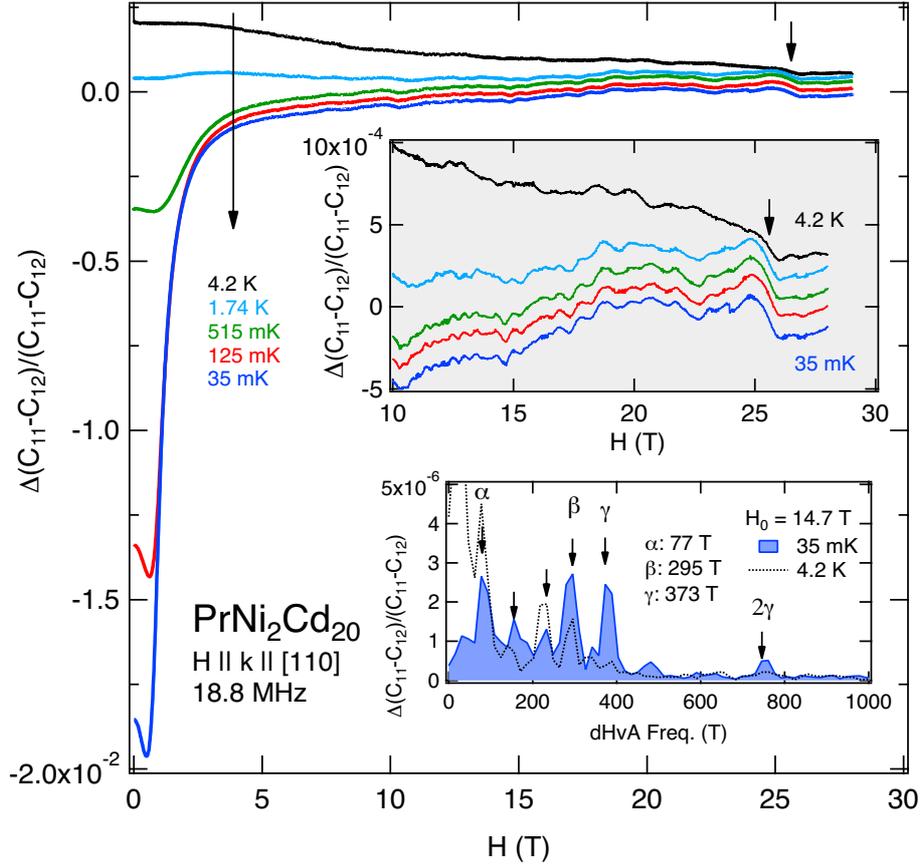

**Figure. 5** Magnetic-field dependence of elastic constant $(C_{11}-C_{12})/2$ of PrNi$_2$Cd$_{20}$. Upper and lower insets show the magnification of the high-magnetic-field region and a Fourier spectrum of the dHvA signals.

In Fig. 5, we show the magnetic-field dependence of the elastic constant $(C_{11}-C_{12})/2$ of PrNi$_2$Cd$_{20}$ up to 28 T at fixed temperatures of 35, 125, and 515 mK and 1.74 and 4.2 K. The elastic constant shows local minima below 1 T, the amount of change in hardening toward higher magnetic fields drastically increasing at lower temperatures. The amount of change in the low-magnetic-field region is consistent with the change in the temperature dependence at zero magnetic field, i.e., the dependence on magnetic field below ~2 T is roughly explained by the quadrupolar susceptibility in the $\Gamma_3$ CEF ground-state model. In Fig. 6, we represent the calculated quadrupolar susceptibility as a function of magnetic field for $H \parallel [110]$ up to 30 T by using two proposed CEF models for Pr$^{3+}$ ($J$ = 4) with $\Gamma_3$ ground state. The $\Gamma_3$–$\Gamma_5$ CEF ground-state model, however, predicts an enhanced softening due to level crossing at approximately 5 T, whereas the $\Gamma_3$–$\Gamma_4$ model demonstrates an enhancement of the softening at approximately 13 T for $H \parallel [110]$. Both calculated results are not consistent with the continuous increase in the elastic constant $(C_{11}-C_{12})/2$ with applied magnetic field as shown in Fig. 5. Thus, such deviation from the CEF calculation also strongly suggests the presence of some field-induced ordered phase in the present compound.

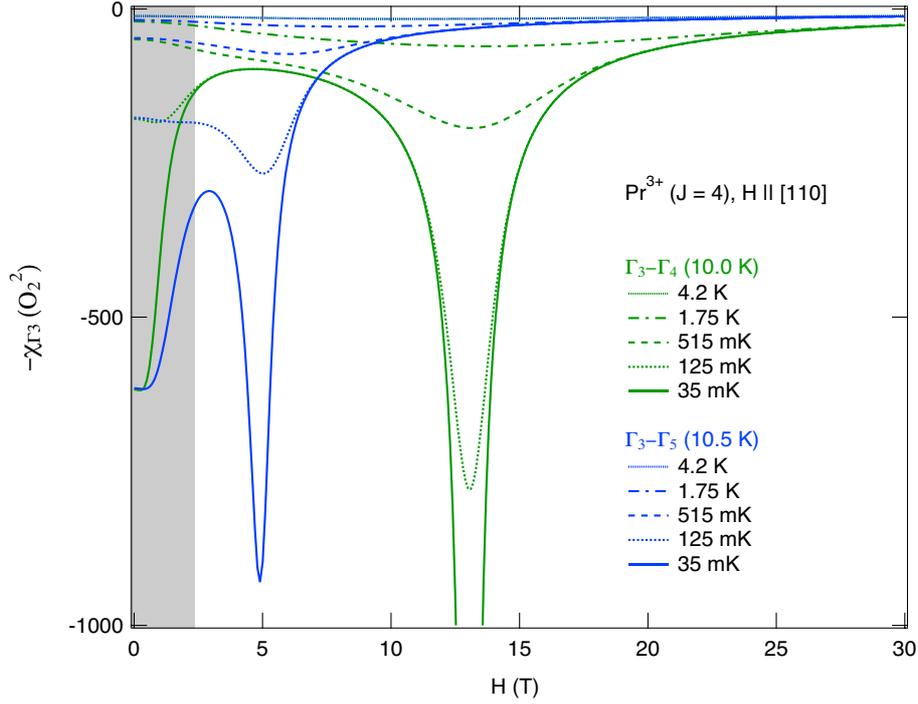

**Figure. 6** Magnetic-field dependence of the calculated quadrupolar susceptibility by using two proposed CEF models for $Pr^{3+}$ ($J = 4$) with $\Gamma_3$ ground state for $H \parallel [110]$ up to 30 T.

On the other hand, acoustic de Haas–van Alphen (dHvA) oscillations [27] are clearly observed in the $(C_{11}-C_{12})/2$ mode as well. By performing the Fourier transformation of the data between 10 and 28 T (median value $H_0 = 14.7$ T), we obtained the dHvA spectrum as shown in the lower inset of Fig. 5, where the dHvA frequencies at 77, 295, and 373 T are estimated from the peaks. These values are distinguished from the expected dHvA signal of excess Cd flux (~500 T) [28]. Further investigation of the angular dependence of the dHvA spectrum and comparison with band-calculation results will be required to understand the band structures and the possibility of the band Jahn–Teller effect on this compound. Another step-like elastic anomaly occurs at approximately 26 T, which is different from the dHvA oscillations and is also not explained by a level crossing of the present CEF-level schemes for the paramagnetic phase. In the magnetic field–temperature phase diagram shown in the inset of Fig. 7, a temperature-independent phase boundary (of a possible high-field phase (HFP)) is drawn at approximately 26 T. A similar phase diagram and temperature-independent phase boundary is also seen in the related compounds $PrRh_2Zn_{20}$ and $PrIr_2Zn_2$, which also have $\Gamma_3$-doublet ground states [22, 23], where the origin is attributed to the dynamical Jahn–Teller effect owing to the coupling between the multipoles and the lattice. It could also be intriguing to consider that the odd-parity multipoles [29, 30] play a role in these high-field anomalies because of the lack of local inversion symmetry at the rare-earth sites. In the low-temperature and low-magnetic-field regions of the phase diagram, we can distinguish several possible field-induced ordered phase, which are indicated as FIOP1 and FIOP2 in Fig. 7. In order to verify these phase boundaries and clarify the origins, it is necessary to investigate the possible magnetic-field anisotropy of the elastic anomalies and conduct their symmetry analysis based on the investigation of the ultrasonic-mode dependence to understand the order parameters of the possible field-induced phases.

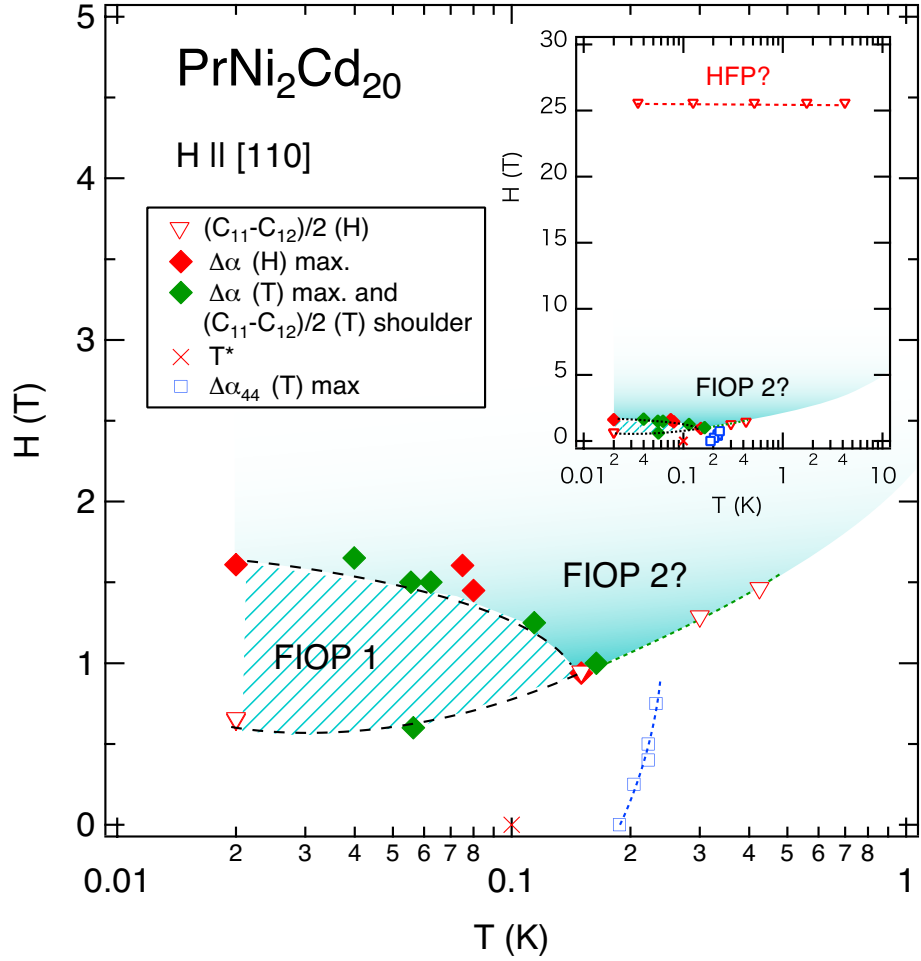

**Figure. 7** Magnetic field–temperature ($H$-$T$) phase diagram of PrNi$_2$Cd$_{20}$ constructed from the present ultrasonic measurements ($H \parallel [110]$). The black-dashed curves show possible phase boundaries of field-induced ordered phases (FIOP1 and FIOP2). The inset shows the same phase diagram for a wider $H$ and $T$ range including the high-magnetic-field anomaly at approximately 25 T. The blue (at ~200 mK in the low-magnetic-field region) and red (at ~25 T) dotted curve are guides to the eye, and not clearly defined as phase boundaries in the present study.

In particular, it is difficult to exactly define the phase boundary in the low-magnetic-field region. This feature evokes a similar $H$-$T$ phase diagram—the Fermi-liquid state without long-range order at low fields—where the antiferroquadrupolar interaction (RKKY interaction) is competing with the Kondo screening ($c$-$f$ hybridization) in, for instance, Ce$_{1-x}$La$_x$B$_6$ [31] and Ce$_3$Pd$_{20}$Si$_6$ [32]. In the present case, a similar scenario can be considered—that the QKE driven by a small but finite $c$-$f$ interaction will be suppressed by a small external magnetic field, which makes the local long-range order evident in intermediate magnetic fields.

## Summary

In summary, we have measured the elastic constant $(C_{11}-C_{12})/2$ of PrNi$_2$Cd$_{20}$ for a wide range of temperatures ($0.02 \leq T \leq 300$ K) and magnetic fields ($0 \leq H \leq 28$ T). From the present study, it is revealed that the CEF ground state of the present compound is the non-magnetic $\Gamma_3$-Kramers doublet and the $\Gamma_4$- or $\Gamma_5$-triplet excited state at ~10 K. In general, a quadrupole order is expected at low $T$ in the cubic $\Gamma_3$-Kramers system. However, a tiny intersite quadrupolar interaction was observed (3.5–4.0 mK), and no clear elastic anomaly in the temperature dependence of the elastic constant for zero magnetic field. These facts imply that the suppression of the long-range order in the low-magnetic-field region of the present system is probably caused by the QKE. On the other hand, several elastic anomalies are found at low temperatures, $T < 0.15$ K, at magnetic fields of $1 \leq H \leq 2$ T. These findings suggest the presence of a magnetic-field-induced local long-range ordered phase.

As the Pr-ionic site, located at the center of the atomic cage, does not have local inversion symmetry (point group $T_d$), this class of compounds belongs to the "parity-mixing system" as well as "cage-structured compounds", both of which have attracted much attention in the research community working on strongly correlated electron systems (SCESs). Therefore, further investigations of the quantum ground state of these systems are essential for the evolution of the research field. In the present work, the acoustic dHvA effect was observed for PrNi$_2$Cd$_{20}$. To verify the possible magnetic-field-induced phase and to understand its origin by help of fermiology studies, further investigation in wider temperature and magnetic field ranges, in particular in the pulsed magnetic field region ($H \leq 60$ T), will also be required.

**Acknowledgements:**

One of the authors (T.Y.) would like to express profound respect for Prof. M. Brian Maple's brilliant achievements that opened the gates to the fertile world of research on cage-structured compounds and the related SCES phenomena originating from research on the filled-skutterudite systems. Research studies at UC San Diego were supported by the US NSF DMR-1206553 and the US DOE under grant no. DE-FG02-04-ER46105. Research studies at Hokkaido University and HZDR were supported by JSPS KAKENHI Grant Nos. 26400342, JP18H04297, JP17K05525, JP18KK0078, JP15H05882, JP15H05885, JP15K21732, and JP19H01832 and the Strategic Young Researcher Overseas Visits Program for Accelerating Brain Circulation. We acknowledge support from the DFG through the Würzburg-Dresden Cluster of Excellence on Complexity and Topology in Quantum Matter, ct.qmat (EXC 2147, project-id 39085490), and from the HLD at HZDR, member of the European Magnetic Field Laboratory (EMFL). This work was also partly supported by the High Field Laboratory for Superconducting Materials, Institute for Materials Research, Tohoku University.